\newcommand\has{\hat{S}}
\newcommand\deq{\delta Q_{rev}}
\newcommand\pa{\partial}
\newcommand\beq{\begin{equation}}
\newcommand\eeq{\end{equation}}
\newcommand\beqnl{\begin{eqnarray}}
\newcommand\beqna{\begin{eqnarray*}}
\newcommand\eeqna{\end{eqnarray*}}
\newcommand\eeqnl{\end{eqnarray}}

 \def\NN{\hbox{\sf I\kern-.13em\hbox{N}}}
 \def\HH{\hbox{\sf I\kern-.13em\hbox{H}}}
 \def\DD{\hbox{\sf I\kern-.13em\hbox{D}}}
 \def\RR{\hbox{\sf I\kern-.14em\hbox{R}}}
 \def\CC{\hbox{\sf I\kern-.44em\hbox{C}}}
 \def\ZZ{{\hbox{\sf Z\kern-.43emZ}}}
 \def\QQ{\hbox{\sf C\kern -.48emQ}}
 \def\Cc{\hbox{\sf C\kern -.47em {\raise .48ex \hbox{$\scriptscriptstyle |$}}
   \kern-.5em {\raise .48ex \hbox{$\scriptscriptstyle |$}} }}
 \def\Qq{\hbox{\sf Q\kern -.57em {\raise .48ex \hbox{$\scriptscriptstyle |$}}
   \kern-.55em {\raise .48ex \hbox{$\scriptscriptstyle |$}} }}

\documentclass{elsart}
\usepackage{amssymb}
\journal{Physics Letters A}
\begin{document}
%\twocolumn
\begin{frontmatter}
%\vskip1pc

\title{Black Hole Thermodynamics in Carath\'eodory's Approach}
\author{F. Belgiorno}
%\footnote{E-mail address: belgiorno@mi.infn.it}}
\address{Dipartimento di Fisica, Universit\`a degli Studi di Milano,\\ 
%\affiliation{Dipartimento di Fisica, Universit\`a degli Studi di Milano,\\ 
Via Celoria 16, 20133 
Milano, Italy}
\ead{belgiorno@mi.infn.it}

\date{\today}

%\maketitle

\begin{abstract}

We show that, in the framework of Carath\'eodory's approach to 
thermodynamics, one can implement black hole thermodynamics 
by realizing that there exixts a quasi-homogeneity symmetry 
of the Pfaffian form $\deq$ representing the infinitesimal heat 
exchanged reversibly by a Kerr-Newman black hole; 
this allow us to calculate readily an integrating factor, and,  
as a consequence, a foliation of 
the thermodynamic manifold can be recovered.

\end{abstract}

\vskip2pc

\end{frontmatter}

\vskip2pc

%\newpage

%\newpage

\section{Introduction}

We consider black hole thermodynamics in the framework of 
Carath\'eodory's approach to thermodynamics, which postulates the 
integrability of the Pfaffian form $\deq$ representing the 
infinitesimal heat exchanged reversibly 
\cite{buchdahl,kestin}. 
The integrability of $\deq$ means that there exist an 
integrating factor $\mu$ and a function $\sigma$ (called 
``empirical entropy'' \cite{buchdahl}) such that 
$\deq = \mu d\sigma$. In standard thermodynamics, the 
integrability of $\deq$ is a consequence of Clausius inequality, 
which ensures that the absolute temperature $T$ is an integrating 
factor for $\deq$, in such a way that $\deq= T dS$, where $S$ 
is the entropy (``metrical entropy'' \cite{buchdahl}). 
The existence of the entropy function is 
part of the second law (the principle of entropy increase for 
thermodynamic processes of closed and thermally insulated systems is  
the other part \cite{kestin}; it is related to the strict inequality, 
which holds in irreversible processes, in Clausius inequality). 
If the integrable Pfaffian form $\deq$ displays a symmetry, in a 
sense to be described in the following,  
it is also possible to calculate explicitly and readily an 
integrating factor by means of elementary tools of differential geometry 
\cite{qotd,belhom,belsym}. 
As a consequence, an explicit construction of the foliation of the 
thermodynamic manifold into disconnected adiabatic hypersurfaces 
satisfying $\deq=0$ is readily calculated.\\   
By postulating a natural  
form for $\deq$in the case of black holes of the Kerr-Newman family, 
we can introduce a notion of temperature and of 
entropy for black holes without referring a priori to 
the laws of black hole mechanics, thanks to the integrability 
of $\deq$ {\sl and} to the presence of 
a quasi-homogeneity symmetry of the Pfaffian form $\deq$. 
Particularly, we can generate a potential  
which is then related to the entropy of the 
black hole. Both the entropy 
and the temperature appear as derived quantities. 
We point out also that the following construction for black holes, 
being made an explicit use both of the integrability condition and 
of the symmetry of $\deq$,  
is of interest also for thermodynamicists.

\section{Pfaffian form and symmetry}
\label{pfaffsymm}

We recall that, for a Kerr-Newman black hole, due to 
the no-hair theorem, the only parameters available are $M,Q,J$ (mass, 
charge and angular momentum);  
for an insulated black hole 
they play the role of conserved charges, 
and are chosen as independent variables in the thermodynamic domain,  
which is assumed to be the 
non-extremal manifold $M^4-M^2 Q^2-J^2>0$ (the extremal 
sub-manifold $M^4-M^2 Q^2-J^2=0$ 
is a boundary of the former, and is temporarily 
not taken into account. 
Some more discussion on 
this topic is found in sect. \ref{thirdlaw}). 
We look for a natural infinitesimal reversible form for the first law. 
Then, we have to find out a Pfaffian form such that the first law 
in its infinitesimal form is implemented, and moreover, such that 
it is integrable (thus, the first part of the second law is ensured). 
In the case of a charged 
rotating system, it would be natural to define 
$\deq = dU+p dV-\phi dq-\omega dl-\mu dN$, where $q$ is the electric charge, 
$l$ is the angular momentum, $\phi$ is the electrostatic potential and 
$\omega$ is the angular velocity. We introduce then the angular velocity 
\beq
\Omega=\frac{J}{M} \frac{1}{2 M^2-Q^2+2 M \sqrt{M^2-Q^2-J^2/M^2}}  
\label{ombh}
\eeq
and the electric potential 
\beq
\Phi=\frac{Q (M+\sqrt{M^2-Q^2-J^2/M^2})}{2 M^2-Q^2+2 M  
\sqrt{M^2-Q^2-J^2/M^2}}  
\label{phibh}
\eeq
of the black hole. 
Both $\Omega$ and $\Phi$ can be assigned on a purely 
geometrical footing, without any a priori knowledge of black hole 
thermodynamics. 
Analogy with standard thermodynamics leads us to 
associate infinitesimal variations of $Q,J$ with work terms $- \Phi dQ$ 
and $- \Omega dJ$;  infinitesimal variations of $M$ can naturally play the 
role of $dU$ in standard thermodynamics, where $U$ is the internal energy. 
We then define
\beq
\deq \equiv dM-\Phi dQ-\Omega dJ. 
\label{dform}
\eeq
Definition (\ref{dform})
is natural, in fact the (rest) mass can be 
identified with (a term of) the internal energy (the rest mass 
of a fluid can be considered as a term of the internal energy in 
standard thermodynamics; see e.g. \cite{guggenheim}); moreover, as seen,  
the work terms appear as standard work terms. The most evident difference 
with the case of a standard system consists in the 
absence, for the black hole case, of the $p dV$ term. This lack 
is associated with the lack of a notion of volume in the black hole case, 
as well known. Even a pressure cannot be defined, and the same is true 
for the particle number $N$.\\
It is stressed that the infinitesimal 
variation $dM-\Phi dQ-\Omega dJ$ is taken along stationary 
black hole solutions of the Kerr-Newman family, because of 
(\ref{ombh}), (\ref{phibh}); this means that the Einstein equations 
are satisfied for each state involved in the aforementioned variation. 
Moreover, these solutions of the Einstein equations 
are considered as black hole equilibrium states,  
to be compared with equilibrium states of standard thermodynamics.\\ 
The Pfaffian form $\deq$ is everywhere non-singular, i.e., there 
is no point of the thermodynamic domain where all the coefficients 
of the differential form vanish (this property holds also for points of 
the extremal boundary). 
It is easy to show that $\deq$ is smooth on the 
non-extremal manifold and is completely integrable, that is, 
it satisfies the condition $\deq \wedge d(\deq)=0$, i.e. 
\beq
-\pa_J \Phi+\pa_Q \Omega+\Phi \pa_M \Omega-\Omega \pa_M \Phi=0.
\label{integr}
\eeq
Being $\deq$ a one-form in three variables, this integrability 
condition is surely non-trivial (it would be trivial in the case 
of two variables). Notice that a different choice for the sign of 
the work terms in (\ref{dform}) would lead to a non-integrable 
Pfaffian form, as it is easy to verify. In the appendix, we 
extend the present considerations to the case where a magnetic 
monopole charge $P$ is allowed. 
It is remarkable that integrability is not postulated, 
but it simply follows from considering infinitesimal 
``on shell'' variations,  
i.e. variations along the aforementioned solutions.\\ 
We can also find an integrating factor by using 
the quasi-homogeneity symmetry of the Pfaffian form 
(\ref{dform}).\footnote{A Pfaffian form 
$
\omega=\sum_{i=1}^n \omega_i(x^1,\ldots,x^n) dx^i
$
is quasi-homogeneous of degree $r\in \RR$ and weights 
$(\alpha_1,\ldots,\alpha_n)\in \RR^n$ if, under the scaling 
$x^1,\ldots,x^n \mapsto 
\lambda^{\alpha_1} x^1,\ldots,\lambda^{\alpha_n} x^n$
one finds $\omega \mapsto \lambda^r \omega$. 
This happens if and only if $\omega_i (x^1,\ldots,x^n)$ is 
quasi-homogeneous of degree $\beta_i=r-\alpha_i$ 
for all $i=1,\ldots,n$, which means that  
$\omega_i (\lambda^{\alpha_1} x^1,\ldots,\lambda^{\alpha_n} x^n)
=\lambda^{r-\alpha_i} \omega_i (x^1,\ldots,x^n)$.}   
In fact, under the quasi-homogeneous transformation 
\cite{anosov} 
(also called ``similarity transformation'' and 
``stretching transformation'' )  
$M\mapsto \lambda^{\alpha} M;\ Q\mapsto \lambda^{\alpha} Q; 
J\mapsto \lambda^{2 \alpha} J$, one obtains 
$\deq\mapsto \lambda^{\alpha} \deq$, i.e., $\deq$ is quasi-homogeneous 
of degree $\alpha$.  
$(\alpha,\alpha,2\alpha)$ are defined to be the weights of 
$M,Q,J$ respectively and they have to be determined. 
Let us define the so-called Euler vector field \cite{anosov}, 
which is infinitesimal generator of the transformation 
\beq
D_{\alpha}\equiv \alpha M \frac{\pa}{\pa M}+
\alpha Q \frac{\pa}{\pa Q}+
2\alpha J  \frac{\pa}{\pa J};
\label{veuler}
\eeq
we have introduced 
above a label $\alpha$ which underlines that $\alpha$ is not yet fixed 
($\{D_{\alpha}\}_{\alpha}$ is a one-parameter family of Euler vector fields); 
let the corresponding Lie derivative be $L_{D_{\alpha}}$; then, 
the quasi-homogeneous transformation is a symmetry for $\deq$ (see e.g. 
\cite{bocharov,cerveau}), in the 
sense that 
\beq
(L_{D_{\alpha}} \deq) \wedge \deq = 0.
\eeq 
In fact, $L_{D_{\alpha}} \deq=\alpha \deq$. 
An integrating factor $f_{\alpha}$ such that 
the form $\deq/f_{\alpha}$ is exact is $f_{\alpha}\equiv i_{D_{\alpha}} \deq
=\deq(D_{\alpha})$. 
For a proof that $f_{\alpha}$ is an integrating factor see Ref. 
\cite{qotd} and, for the homogeneous case, see  
e.g. Ref. \cite{cerveau} and also 
Ref. \cite{belhom}, 
where an application to ordinary thermodynamics can be found. 
In our case, one obtains 
$f_{\alpha}=\alpha (M-\Phi Q-2 \Omega J)$,  
explicitly $f_{\alpha}=\alpha \sqrt{M^2-Q^2-J^2/M^2}$, which is not 
identically vanishing, thus $D_{\alpha}$ is associated with a transversal 
(or non-trivial) 
symmetry \cite{qotd} (i.e., $D_{\alpha}$ does not belong 
to the distribution of codimension one associated with the kernel of 
$\deq$). We remark that the integrating factor $f_{\alpha}$ is 
proportional to the horizon coordinate ${\it c}$ introduced by B.Carter 
in \cite{cartergeo}. Then, $f_{\alpha}$ is constant on the horizon. 
 
\section{Foliation of the thermodynamic manifold}
\label{thermfol}

Frobenius theorem for the Pfaffian form $\deq$ on the 
non-extremal manifold 
can be invoked and a foliation of 
the non-extremal manifold can be generated thanks to the 
integrability property (\ref{integr}). 
The non-extremal thermodynamic space is foliated by the submanifolds 
(of codimension one) which are solutions of the Pfaffian equation 
$\deq=0$. 
The leaves of the foliation of codimension one 
are surfaces where the potential associated with $\deq/f_{\alpha}$ is 
constant. Let $(M_{0},Q_{0},J_{0})$ be a reference state and 
$\Gamma$ be any path connecting the reference state to the 
state $(M,Q,J)$ of interest. By choosing  e.g. a rectangular path 
$(M_{0},Q_{0},J_{0})\to (M,Q_{0},J_{0})\to (M,Q,J_{0}) 
\to (M,Q,J)$ contained in the non-extremal manifold, one finds 
\beqnl
\hat{S}_{\alpha}(M,Q,J)-\hat{S}_{\alpha}(M_{0},Q_{0},J_{0})
&\equiv& \int_{\Gamma} \frac{\delta Q_{rev}}{f_{\alpha}}\cr
&=&
\frac{1}{2\alpha} \log \left(
\frac{M^2 b^2(M,Q,J)+J^2/M^2}{M_{0}^2 
b^2(M_{0},Q_{0},J_{0}) +J_{0}^2/M_{0}^2}\right),
\eeqnl
where $b(M,Q,J)\equiv (1+\sqrt{1-Q^2/M^2-J^2/M^4})$. 
The argument of the logarithm is proportional 
to the black hole area 
$A=4 \pi (M^2 b^2(M,Q,J)+J^2/M^2)$. We have generated a foliation 
of the parameter space of Kerr-Newman black holes. The 
leaves are the surfaces $A=$ const., as expected, but we cannot yet 
determine the so-called metrical entropy \cite{buchdahl} 
in the case of black holes.\\ 
 We now introduce an assumption which requires some discussion.  
The above 
procedure is a generalization, discussed in Ref. \cite{qotd}, 
of the procedure one can develop  
for standard thermodynamics \cite{belhom}. 
In the 
case of standard thermodynamics of homogeneous systems, 
the Pfaffian form $\deq=dU+p dV-\mu dN$ 
in Gibbsian variables $(U,V,N)$ is homogeneous  
(for the definition of homogeneous differential form see 
e.g. Ref. \cite{cerveau}). The generator of the 
symmetry is the ``Liouville'' operator 
%$Y=U \frac{\pa}{\pa U}+V \frac{\pa}{\pa V}+N \frac{\pa}{\pa N}$ 
$Y=U \pa_U+V \pa_V+N \pa_N$ 
and the integrating factor is $\deq(Y)=U+p V-\mu N$.  
For a standard thermodynamic 
system one finds that $d\has\equiv \deq/f =dS/S$, where $S$ is an extensive 
function which coincides with the metrical entropy of the 
system and corresponds to the fundamental relation in the entropy 
representation \cite{belhom}. This deduction is corroborated by 
appealing to the homogeneity of $S$ in Gibbs' approach, which allows to 
find $T S=U+p V-\mu N=\deq(Y)$, i.e., the integrating factor 
coincides with $T S$. 
We proceed by analogy 
with the formalism of thermodynamics just sketched, which means that 
{\sl we assume that the metrical entropy is the unique
\footnote{Uniqueness holds within a 
multiplicative constant. See \cite{qotd}.} quasi-homogeneous function $S$ 
of degree one which satisfies $d\has\equiv\deq/f = dS/S$}.  
We refer to \cite{qotd} for 
a proof that such an $S$ exists and is unique. 
This assumption about the role of $S$ such that $\deq/f = dS/S$ 
is correct both in the case of standard thermodynamics in the Gibbs 
space \cite{belhom} and in the case of black hole thermodynamics, 
and the latter case is analyzed in the following.\\ 
The potential $S_{\alpha}$ such that $d\has_{\alpha}=dS_{\alpha}/S_{\alpha}$ 
for the black hole case is
\beq
S_{\alpha}=c_{\alpha} A^{1/{2\alpha}}
\label{salfa}
\eeq
where $c_{\alpha}$ is an undetermined constant. We  have a one-parameter 
family of possible metrical entropies (and also fundamental relations in 
the entropy representation) which satisfy  
$D_{\alpha} S_{\alpha}=S_{\alpha}$, in analogy with 
$Y S=S$ of standard thermodynamics.  
Our result (\ref{salfa}) agrees with the result contained in 
Ref. \cite{gour} but we work in a more general 
framework where no reference to the laws of black hole 
mechanics is made [notice also that in our expression for 
$S$ no additive constant appears, due to quasi-homogeneity symmetry].  
There is still an ambiguity due to the undetermined value of 
$\alpha$, which means that we know the ratio between the weights 
of $M,Q,J$ but not yet the weights themselves. Notice that 
this ambiguity does not 
occur in the case of standard thermodynamics, where 
the weights of $(U,V,N)$ are known and they are all equal to one.  
We recall that 
the metrical entropy is assumed to belong to the one-parameter 
family $\{ S_{\alpha} \}_{\alpha}$. 
For each $\alpha$ the temperature is 
$T_{\alpha}=(\pa S_{\alpha}/\pa M)^{-1}$ and it is a quasi-homogeneous 
function of degree $\alpha-1$ and weights $(\alpha,\alpha,2\alpha)$. 
It is useful to realize that 
\beqnl
S_{\alpha}&=&\frac{c_{\alpha}}{(c_{1/2})^{1/2\alpha}} 
(c_{1/2} A)^{1/{2\alpha}}=
\frac{c_{\alpha}}{(c_{1/2})^{1/2\alpha}} (S_{1/2})^{1/{2\alpha}}\\
T_{\alpha}&=&2\alpha 
\frac{(c_{1/2})^{1/2\alpha}}{c_{\alpha}} (S_{1/2})^{1-1/2\alpha} 
T_{1/2}.
\eeqnl
For any $\alpha$ one gets 
$T_{\alpha} dS_{\alpha}=dM-\Phi dQ-\Omega dJ$. The 
black hole area is known to be a superadditive function of 
$M,Q,J$.  
Superadditivity of the entropy, which plays a fundamental role 
when one considers the merging of two black holes, does not 
fix $\alpha$.
\footnote{$0<\alpha \leq 1/2$ is a sufficient condition 
for preserving superadditivity.} 
Cf. also 
\cite{gour}  for a discussion 
concerning an analogous ambiguity 
(our $1/2\alpha$ is $\gamma$ therein). 
The Hawking effect is necessary in order to give us an actual 
thermodynamic meaning to our calculation;  it also fixes 
$\alpha$, in fact, in order to identify  
the temperature of the black hole with the 
Hawking one it is mandatory to choose $\alpha = 1/2$. 
There is also a multiplicative constant (namely, $c_{1/2}$) which has to be 
determined. By comparison with 
the Hawking effect, one finds that $c_{1/2}=1/4$. The above ambiguity 
can also be resolved phenomenologically, in perfect agreement with 
the phenomenological nature of a thermodynamic approach \cite{qotd};   
one should determine $M,Q,J$ and then plot $T(M,Q,J)$ from measurements 
of the temperature. $\alpha = 1/2$ (and also $c_{1/2}=1/4$) should come out 
again.

%\subsection{Generalized Gibbs-Duhem equation for black holes}

It is remarkable that, as a consequence of the quasi-homogeneity 
of black hole entropy, one gets the following 
generalized Gibbs-Duhem equation, which is analogous to the 
Gibbs-Duhem equation of standard thermodynamics:
\beq
M^2 d\left(\frac{1}{2 M T}\right) 
- Q^2 d\left(\frac{\Phi}{2 Q T}\right)
-J d\left(\frac{\Omega}{T}\right)=0.
\label{gibh}
\eeq
This follows from $S=i_D (\deq/T)$ and from 
$dS=\deq/T$, where $D\equiv D_{1/2}$. 
It is easy to show that this generalized Gibbs-Duhem equation 
corresponds to 
\beq
-i_D d(\frac{\deq}{T})=0.
\label{giduh}
\eeq 
In fact, $dS=-i_D d (\deq/T) + L_D (\deq/T)=
-i_D d (\deq/T)+\deq/T$; 
from the latter equality and from $dS=\deq/T$ equation (\ref{giduh}) 
follows. A simple rearrangement of the terms one obtains by 
making explicit (\ref{giduh}) gives then (\ref{gibh}). 
See \cite{qotd} for a general setting and \cite{qotn} for the case of 
standard thermodynamics.

%\subsection{Comparison with black hole  mechanics}

Contrarily to the 
naive expectation, the laws of black hole mechanics  
give no unique hints 
about the value of $\alpha$, they don't fix uniquely the metrical 
entropy and the absolute temperature of the black hole. 
For any $\alpha$ one gets 
$T_{\alpha} dS_{\alpha}=dM-\Phi dQ-\Omega dJ$, to be compared 
with the differential form of the first law. 
Moreover, one finds that $f_{\alpha}=T_{\alpha} S_{\alpha}$, which   
implies $\alpha (M-\Phi Q-2 \Omega J)=
T_{\alpha} S_{\alpha}=2\alpha T_{1/2} S_{1/2}$. 
By comparison with the first law in the finite form 
one realizes that $T_{1/2}=k/(8 \pi c_{1/2})$. 
The choice of a generic $\alpha$ is equivalent to the the substitutions
$A \mapsto {\bar A}_{\alpha}$ and $k \mapsto {\bar k}_{\alpha}$, where 
${\bar A}_{\alpha} = A^{1/{2\alpha}}$ and 
${\bar k}_{\alpha} =2\alpha k/A^{1/{2\alpha}-1}$, 
which implement both 
the differential form and the finite form of the first law (the  
latter appears as 
${\bar k}_{\alpha} {\bar A}_{\alpha}=
8\pi \alpha (M-\Phi Q-2 \Omega J)$ 
which is equivalent to the well-known one).  
Notice that ${\bar k}_{\alpha}$ is constant on the horizon, thus the 
zeroth law of black hole mechanics is not sufficient in order to 
select $\alpha=1/2$.

\section{The extremal boundary}
\label{thirdlaw}

The extremal submanifold is very problematic. 
It is easy to show that 
$\deq=0$  on the extremal submanifold, i.e.  
the extremal submanifold is still an integral submanifold of the 
Pfaffian form \cite{belbh}. 
Nevertheless, there is an important property 
which fails in the case of states belonging to the extremal 
submanifold. In fact, given a point of the extremal submanifold, 
there exist two kinds of adiabatic paths having the given state as 
initial point. One is a path lying on the extremal submanifold, 
the other is an ``isoareal'' path, i.e. a path starting from the 
extremal submanifold and reaching non-extremal states 
each of which has the same area as the initial extremal state 
\cite{belbh}. In absence of the latter class of solutions, 
the extremal states would represent a leaf of a foliation, thus 
they would be adiabatically disconnected from the non-extremal 
states. Instead, their presence can jeopardize the second law 
of thermodynamics. 
A detailed discussion 
of this topic and of the third law in black hole thermodynamics 
is the subject of Ref. \cite{belbh}. See also \cite{belg3} for 
the case of standard thermodynamics.

\section{Conclusions}

The approach to black hole thermodynamics by means of 
Pfaffian forms we have discussed (Carath\'eodory's formalism) 
represents a further corroboration of the fact that black hole 
thermodynamics is a form of thermodynamics, even if 
to large extent exceptional.  Quasi-homogeneity symmetry of $\deq$ plays an 
important role in allowing to calculate an integrating factor and, then, 
to generate a thermodynamic potential depending on a undetermined 
parameter $\alpha$ which can nevertheless be fixed phenomenologically, 
as seen.  
Notice that our approach can be 
extended in a straightforward way to  
KN-AdS black holes \cite{caldarelli}. Also in this case, there 
is a quasi-homogeneity structure in the Pfaffian form, as it can 
be easily realized. It is remarkable that quasi-homogeneity is a 
symmetry occurring also in the thermodynamics of other self-gravitating 
systems, like non-relativistic fermionic matter and 
self-gravitating radiation. Independent thermodynamic variables and their 
weights  change \footnote{One has $(U,V,N)$ as independent variables 
in the fermionic matter case, and $(U,V)$ in the self-gravitating 
radiation case.}, but a quasi-homogeneous symmetry appears again. 
See \cite{qotd} on the latter topic.   
The corresponding lack of homogeneity can be related to the 
purely attractive nature of gravity. 

This kind of thermodynamic approach can be insightful also from the 
point of view of a more general discussion concerning horizon 
thermodynamics. The availability of a meaningful $\deq$ allows to 
discriminate between cases where there is a 
complete thermodynamic structure at hand, which can be associated with 
the Einstein equations (e.g. because $\deq$ is integrable ``on shell'') 
and cases where Hawking temperature seems to be 
related simply to kinematics.  
Cf. \cite{visser}. 
The case of black holes belongs to the former class, 
a full thermodynamic structure exists and 
the role of the equations of General Relativity in ensuring the 
laws of thermodynamics, enhanced e.g. in Ref. \cite{visser}, is 
corroborated in this framework. Notice that, 
from this point of view, 
the behavior of General Relativity is, to some 
extent, intermediate with respect to a macroscopic phenomenological 
approach, like classical thermodynamics, and a microscopic 
approach, like statistical mechanics. 
A macroscopic (``thermodynamic'') point of view is adopted 
in treating variables like $M,Q,J$; on the other hand,  field equations 
furnish $\Phi,\Omega$ and ensure an integrability 
condition which, for standard systems, should be an outcome  
of statistical mechanics [statistical mechanics  
should calculate the analytic form of the functions $\Phi,\Omega$ 
(they are phenomenological interpolations for 
thermodynamics); moreover, it should 
justify an integrability condition which 
is only a postulate in standard thermodynamics; statistical mechanics should 
allow to determine both the metrical entropy and 
the weights of the variables $M,Q,J$ \cite{qotd}]. As far as a 
``cosmological horizon'' like De Sitter (one parameter) is concerned, 
it corresponds to a solution of General Relativity and $\deq$ can be given 
(but a too tight thermodynamic space does not allow a non-trivial  
integrability condition). 
On the other hand, 
in the ``acceleration horizon'' case (like e.g. 
Rindler case) there is a too poor thermodynamic structure, 
in the sense that there is no first law. The same comment holds true 
in the case of ``acoustic horizons'' (no first law is known).

\appendix

\section{Magnetically charged black holes}

An extension is represented by a rotating charged black hole:
\beq
\deq=dM-\Phi dQ-\Psi dP-\Omega dJ,
\eeq
where  $P$ is the magnetic monopole 
charge and
\beqnl
&&\Omega=\frac{J}{M} 
\frac{1}{2 M^2-Q^2-P^2+2 M \sqrt{M^2-Q^2-P^2-J^2/M^2}},\\  
&&\Phi=\frac{Q (M+\sqrt{M^2-Q^2-P^2-J^2/M^2})}{2 M^2-Q^2-P^2+2 M  
\sqrt{M^2-Q^2-P^2-J^2/M^2}},\\  
&&\Psi=\frac{P (M+\sqrt{M^2-Q^2-P^2-J^2/M^2})}{2 M^2-Q^2-P^2+2 M  
\sqrt{M^2-Q^2-P^2-J^2/M^2}}.  
\eeqnl
Notice that $\deq$ is ``on shell'', i.e., it is taken along (stationary) 
solutions of General Relativity.  
$\deq\wedge d(\deq)=0$ corresponds to four 
integrability conditions:
\beqnl
&&l_{MQP}\equiv 
-\pa_P \Phi+\pa_Q \Psi+\Phi \pa_M \Psi-\Psi \pa_M \Phi=0,\\
&&l_{MQJ}\equiv
-\pa_J \Phi+\pa_Q \Omega+\Phi \pa_M \Omega-\Omega \pa_M \Phi=0,\\
&&l_{MPJ}\equiv
-\pa_J \Psi+\pa_P \Omega+\Psi \pa_M \Omega-\Omega \pa_M \Psi=0,\\
&&l_{QPJ}\equiv 
\Phi (\pa_J \Psi-\pa_P \Omega)-\Psi (\pa_J \Phi -\pa_Q \Omega)
+\Omega (\pa_P \Phi-\pa_Q \Psi)=0.
\eeqnl
Even in this case, the integrability is verified and 
$\deq$ is a quasi-homogeneous Pfaffian form. The Euler vector field 
is given by 
\beq
F=\frac{1}{2} M \frac{\pa }{\pa M}+
\frac{1}{2} Q \frac{\pa }{\pa Q}+
\frac{1}{2} P \frac{\pa }{\pa P}+J \frac{\pa }{\pa J} .
\eeq
The integrating factor is easily found, and, again the area law can be 
obtained.

\end{document}